\documentclass[12pt]{article}
\usepackage[margin=1in]{geometry}
\usepackage[utf8]{inputenc}
\usepackage{amsmath}
 \usepackage{setspace}

\usepackage{authblk}
\usepackage{graphicx}
\usepackage{comment}
\usepackage{algorithmic}
\usepackage{enumitem}
\usepackage[english]{babel}
\usepackage[nottoc]{tocbibind}
\usepackage{url}
\usepackage{color}

\providecommand{\keywords}[1]
{
  \small	
  \textbf{\textit{Keywords---}} #1
}

\title{Linking open source code commits and MOOC grades to evaluate massive online open peer review}

\author[1]{Siruo Wang}
\author[1]{Leah R. Jager}
\author[2]{Kai Kammers}
\author[3]{Aboozar Hadavand}
\author[1]{Jeffrey T. Leek}

\affil[1]{Department of Biostatistics, Johns Hopkins Bloomberg School of Public Health, Baltimore, MD, USA}
\affil[2]{Division of Biostatistics and Bioinformatics, Department of Oncology, Sidney Kimmel Comprehensive Cancer Center, Johns Hopkins University School of Medicine, Baltimore, MD, USA}
\affil[3]{School of Computational Sciences, Minerva Schools at Keck Graduate Institute, San Francisco, CA, USA}
\date{}

\doublespace
\begin{document}
\maketitle

\begin{abstract}
Massive Open Online Courses (MOOCs) have been used by students as a low-cost and low-touch educational credential in a variety of fields. Understanding the grading mechanisms behind these course assignments is important for evaluating MOOC credentials. A common approach to grading free-response assignments is massive scale peer-review, specially used for assignments that are not easy to grade programmatically. It is difficult to assess these approaches since the responses typically require human evaluation. Here we link data from public code repositories on GitHub and course grades for a large massive-online open course to study the dynamics of massive scale peer review. This has important implications for understanding the dynamics of difficult to grade assignments. Since the research was not hypothesis driven, we described the results in an exploratory framework. We find three distinct clusters of repeated peer-review submissions and use these clusters to study how grades change in response to changes in code submissions. Our exploration also leads to an important observation that massive scale peer-review scores are highly variable, increase, on average, with repeated submissions, and changes in scores are not closely tied to the code changes that form the basis for the re-submissions.

\end{abstract}


\section{Introduction}
\label{sec:introduction}

Massive Open Online Courses (MOOCs) are free educational online courses available for a large number of students to participate in. MOOC platforms contain thousands of courses taught by instructors from
universities and institutions around the world \cite{yuan2013moocs,clow2013MOOCs}. Components of MOOC courses may include online recorded video lectures, graded assignments, and discussion forums open to the community, and the goal of the MOOC platform is to provide a low-cost, open-access, and short time training platform to enhance education in specific areas. The courses provided in the MOOC platform often manage student learning process by designing a final project at the end of course content to test student learning ability. Technical MOOC platforms frequently use Git and GitHub \cite{blischak2016quick} as tools for students to upload their final projects for completion of MOOC courses. Git is a version control system to manage changes to files and projects over time, and GitHub is an open-source website that hosts online Git repositories (i.e. repository is a folder that contains files and sub-folders through commits). Using GitHub, students can easily add, delete, and keep track of changes to files and also have a convenient way to share and collaborate with other students in the course. 

Among the many courses taught through the MOOC platform, data analysis is one of the popular areas as data analysis skills are often required when students are looking for job openings \cite{romero2017educational}. In this paper, we focus on one entry-level data analysis MOOC course -- "Getting and Cleaning Data" (\url{https://www.coursera.org/learn/data-cleaning}) provided in Coursera, that contains a fundamental but crucial part of data analysis: learning to access and clean data. This course is a great representation of MOOC courses using peer-reviewed schema to evaluate student class performance. Students who enroll in this MOOC and wish to receive a course completion certificate are required to complete a final project. Students are required to perform two actions to complete the final assignment -- \textit{commit} code on GitHub and \textit{submit} a repository link in the MOOC. In detail, the student first needs to create a repository on GitHub and \textit{commit} code for the course final project to this repository. The course instruction requires three files committed to the GitHub repository -- a "run\_analysis.R" file to demonstrate student ability to collect, work with, and clean a wearable computing data set, a "readme.md" file to describe how the R script works, and a "codebook.md" file to describe the variables. Then student must copy the GitHub repository link that includes the final project and \textit{submit} it in the MOOC platform. In this way, peer graders in the course can access and evaluate the student code performance on GitHub.

Although students in the "Getting and Cleaning Data" MOOC share the same project goal of preparing a tidy data set for further data analysis, they manage GitHub code \textit{commit} and MOOC link \textit{submit} behaviors in different ways. For example, some students prefer to submit their repository link in MOOC one time after committing their code on GitHub. Other students prefer to submit their repository link multiple times while they are making several commits on GitHub.

The "Getting and Cleaning Data" Coursera course is a representative of using peer-review schema in MOOCs since the course final project includes both coding logic and description of variables used in codes that require human evaluation. Peer-reviewed assignments are difficult to grade programmatically by definition, and thus we we focus on exploring potential relationships between student code submissions on GitHub and peer-reviwed grades in this MOOC. This has important implications for understanding the dynamics of difficult to grade assignments. When the MOOC platform receives a student submission, it will randomly assign this submission to multiple peers for grading, and the final peer grade for this submission is the average grade received from peer graders. There is no easy way to evaluate how well the massive scale peer-review system works for a course because an outside assessment of the peer reviews would require massive human effort. Instead of re-evaluating each submitted project, we consider analyzing information from public code commits on GitHub to better understand the MOOC peer-review process. Our analysis did not focus on improving MOOC grading evaluation system and the validity of crowdsourcing \cite{zheng2017truth}. Rather we focus on the relationship between crowdsourced grades as assigned by averaging multiple reviewers scores on Coursera and student GitHub code commit histories. 
In this paper, we look at student code commit behaviors on GitHub for completing the MOOC course to explore the main sources of variations in the massive scale peer-review scores of an entry-level data analysis course.

\section{Data description Overview}
\label{sec:data_description}

\begin{figure}
\centering
\includegraphics[width=\linewidth]{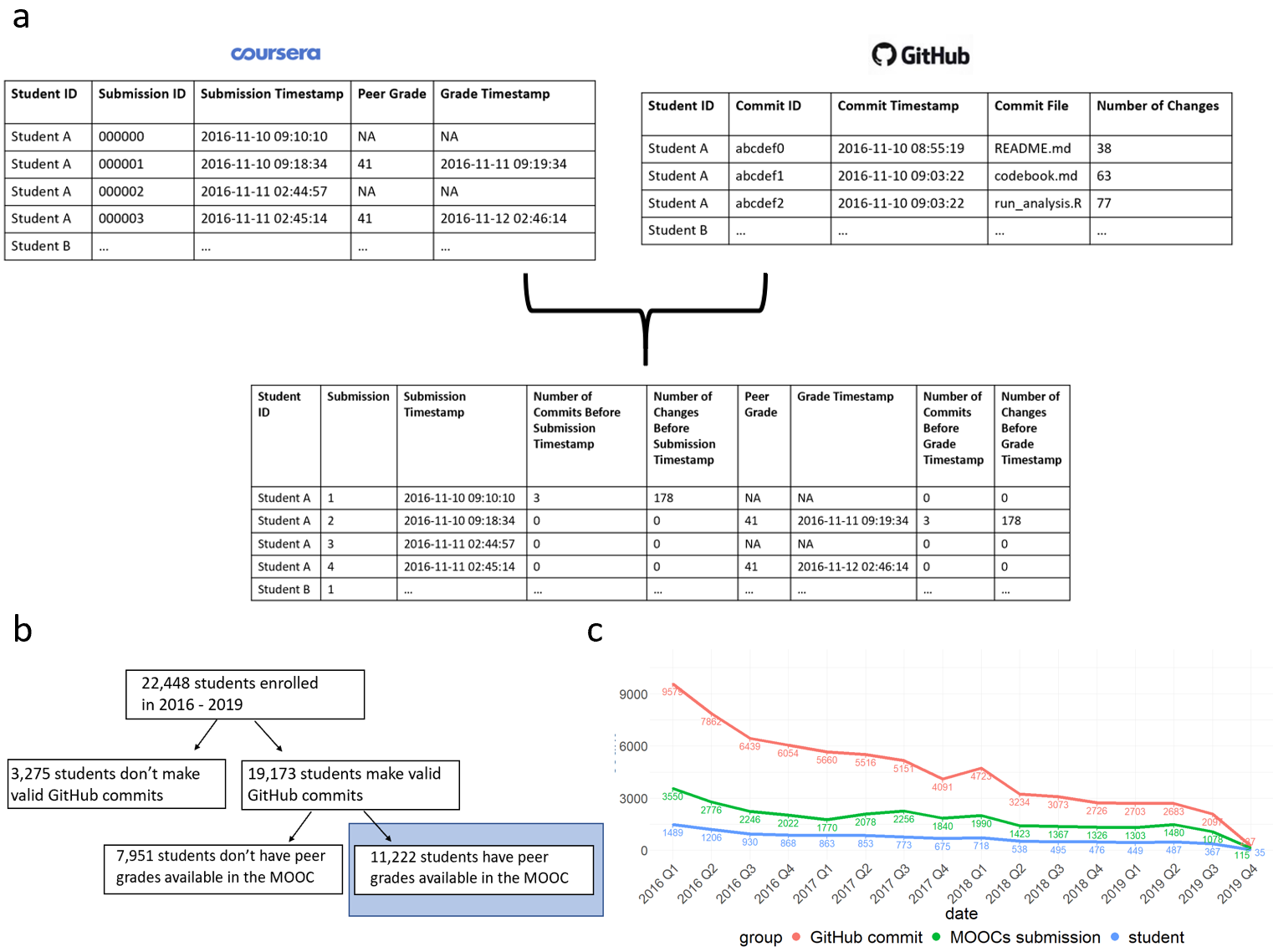}
\caption{\textbf{Data flowchart and student submissions and commits distribution over time.} We show panel (a) in the top left a sample data table collected in MOOC, in the top right a sample data table collected on GitHub, and in the bottom a combined data table with linked unique student IDs and UTC timestamps. We show panel (b) a data description flowchart with blue color to highlight the target population in this study. We show panel (c) a plot on the number of students enrolled in MOOC course (blue color), the number of student submissions in MOOC (green color), and the number of student code commits on GitHub (red color) over time. On the x-axis are the date in quarter yearly periods, and on the y-axis are the counts.}
\label{fig:figure_1}
\end{figure}

The "Getting and Cleaning Data" MOOC course was created online in 2016. We collected data on MOOC submissions, peer grades, and GitHub code commit changes for students who enrolled in this course from January 2016 to November 2019 (IRB approval was obtained from our university IRB office). Within the MOOC database, each student is assigned a unique student ID once registered for the course. There is also a unique student submission ID generated for each assignment submission, along with a submission timestamp. In this course, there are no limits on how many MOOC submissions a student can make before the final project deadline. However, we can use the unique student IDs to match all submission IDs associated with this student. We then use the unique submission IDs to look up student submission information in the MOOC data. For each unique submission ID, we collect information on the submission timestamp, the average peer grade assigned to this submission, and the grade timestamp indicating when this submission is graded. Figure \ref{fig:figure_1}, Panel (a) shows a sample data table from the Coursera MOOC database.  In the first column of the sample data table, we organize enrolled students using their unique student IDs. In the second column, we use the unique student IDs to find all unique submission IDs linked to the students. In the third, fourth, and fifth columns, we use the unique submission IDs to find the submission timestamp, peer grade, and grade timestamp linked to each submission.

To complete this course project, the student is required to create a repository on GitHub, make code commits to the GitHub repository, and submit the repository link through the MOOC platform. When a student pushes code changes to files on GitHub, a unique commit ID and a commit timestamp are generated on GitHub. Using the unique commit ID, we can further determine which of the three files (i.e. "readme.md", "codebook.md", or "run\_analysis.R") that the student pushed changes to and how many code changes were made to the file. There are no limits on the number of commits or the number of changes that student can push on GitHub. The repository links submitted through the MOOC are then used to connect unique students IDs from the MOOC to unique commit IDs generated on GitHub. For each unique student who submits in MOOC, we pull GitHub commit history using the GitHub API to collect student code commit information. Figure \ref{fig:figure_1}, Panel (a) also shows a sample data table from GitHub. In the first column of the table, we label the unique student IDs collected from the MOOC. In the second column, we find all unique commit IDs that link to each student ID. In the third, fourth, and fifth columns, we use the unique commit IDs to find the commit timestamp, commit file, and the number of changes made to the specific file.

Both the MOOC and GitHub platforms use the same strategy that involves the date-time-offset pattern. In our data, we collect three important timestamps -- the submission timestamp that indicates when a student submits the MOOC, the grade timestamp that indicates when MOOC assigns the final peer grade, and the commit timestamp that indicates when a student commits on GitHub. To trace student submissions in MOOC and code commits on GitHub over time, we rank each student's submissions by submission timestamps in the MOOC data table and rank each student's commits by commit timestamps in the GitHub data table. Because both the MOOC and GitHub data tables share the same unique student IDs, we use this to combine the two tables into one large table as shown in the bottom of figure \ref{fig:figure_1} panel (a). In the combined table, we further compare student submission and commit activities using the ranked timestamps. In the combined table, we summarize student commit activities to all three files in between student submission timestamps and grade timestamps. For example, the numbers shown in the third and fourth columns of the commit table indicate that there are three commits and 178 changes that student A has made on GitHub before the first submission in MOOC at 2016-11-10 09:10:10. Student A then does not make any commits or changes from the second to the fourth submissions. The numbers shown in the eighth and ninth columns indicate that there are three commits and 178 changes that student A has made on GitHub before the first peer grade assigned at 2016-11-11 09:19:34. Student A then does not make any commits or changes have made before receiving the second peer grade at 2016-11-12 02:46:14.

In the MOOC database from January 2016 to November 2019, there are 22,448 unique students enrolled in the course, and these 22,448 students make 51,648 submissions to complete the final project in MOOC. We can keep track of students and the submissions they make using unique student IDs and submissions IDs generated in this MOOC. We show the structure of our data in figure \ref{fig:figure_1} panel (b). Among the enrolled 22,448 students, not every student follows the final project instruction to commit the three required files -- "readme.md", "codebook.md", and "run\_analysis.R" files on GitHub. To identify which student follows the instruction, we develop a system using GitHub API for automatically retrieving student code commit history and the data analysis R script on GitHub using the GitHub repository link students provide in their submissions. However, not all GitHub repository links provided in the MOOC are valid links. During the process of scraping student commit history from GitHub, we find the following cases that we cannot retrieve valid data from GitHub: (1) no GitHub repository link is provided for some submissions, (2) the provided GitHub repository no longer exists, (3) no R script found in the GitHub repository, and (4) R script found but is not named as instructed for the project. Considering these exceptions, we exclude 3,275 students who provide repository links in the above four categories. In the remaining 19,173 students who make valid GitHub commits, we only consider students who receive peer grades in MOOC for our study. Although a student can make many submissions, not every submission gets assigned to a peer grade. It is also possible that none of the submissions for a student has a peer grade. This is because the peer-reviewed grads are only assigned to students who make payments to the MOOC course, though non-payers are also allowed to take the course and are encouraged to submit their final projects to practice data analysis skills. Since we will associate student submissions and code commit patterns to peer grades, we only focus on the desired subset of 11,222 students who make valid code commits on GitHub and also receive peer grades for at least one of their submission in MOOC. We use a blue box in figure \ref{fig:figure_1} panel (b) to highlight this subset of students for our study.

For our target population of 11,222 students who enrolled in the course and completed their last submissions from the first quarter in 2016 to the fourth quarter in 2019, in figure \ref{fig:figure_1} panel (c), we show the number of students enrolled in the MOOC course (blue color), the number of submissions in MOOC (green color), and the number of commits on GitHub (red color) in quarterly year periods. When retrieving the commit numbers, we count all code commits for the three required files for the final project. We observe that the overall trend of enrolled students, student submissions, and student code commits are decreasing. On average, one student makes two submissions in MOOC and six commits on GitHub, and the average numbers keep relatively the same across quarter yearly periods from 2016 to 2019.

The remainder of the paper is organized as follows. In section \ref{sec:submission_commit}, we observe student submission patterns in MOOC and study student code commit patterns on GitHub happening in between MOOC submissions. In section \ref{sec:submission_grade}, we trace student peer-reviewed grades over submissions in MOOC and study the relationship between re-submissions and peer grade increases. In section \ref{sec:commit_grade}, we further associate the variability in student peer grades to the number of code changes on GitHub. We conclude in Section \ref{sec:discuss}.

\section{Patterns for MOOC submissions and GitHub commits}
\label{sec:submission_commit}

\begin{figure}
\centering
\includegraphics[width=\linewidth]{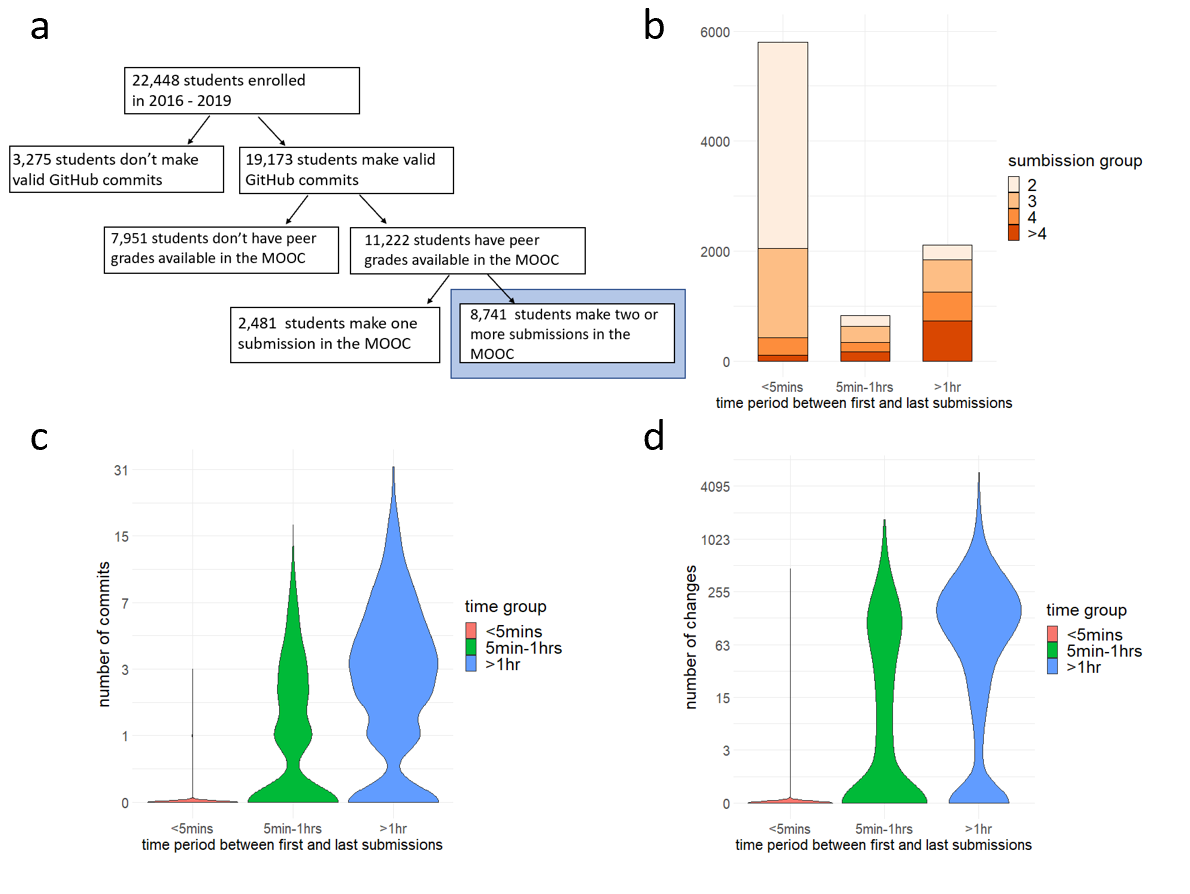}
\caption{\textbf{Student MOOC submissions and GitHub commits.} We show panel (a) the data description flowchart with blue color to highlight the desired population in this study. For panel (b), (c), and (d), on the x-axis are the three time-period groups between student first and last submissions. On the y-axis are (b) the number of students in each time-period group with colors to highlight the number of submissions, (c) the number of commits on GitHub, and (d) the number of changes on GitHub happened within each of the three time-period.}
\label{fig:figure_2}
\end{figure}

In Section \ref{sec:data_description}, we introduce a subset of 11,222 students who enrolled in the MOOC course, provided valid GitHub repository links, and received at least one peer grade among submissions. Within this population, we observe that students make between 1 and 45 submissions, and these 11,222 students have submitted their final projects a total of 28,620 times in the MOOC. Although there is a large range of submission numbers, students mostly submit one or two times and rarely submit four times or more:  22\% of students submit one time, 38\% of students submit two times, 22\% of students submit three times, and 18\% of students submit four times or more. Among the 11,222 students, 2,481 students make only one submission in MOOC. Here we focus on the remaining 8,741 students who make two or more submissions in this MOOC. These two subsets are shown in the data structure flowchart in figure \ref{fig:figure_2} panel (a), which is an extension of the flowchart described in figure \ref{fig:figure_1}(a). The subset of 8,741 students for our analysis is also highlighted with blue color in figure \ref{fig:figure_2}(a).

For the target population of 8,741 students, we rank each student submissions by submission timestamps and calculate the time-periods between student first and last submission. We observe that more than half of the population (5,803 students) complete all of their submissions within five minutes. Based on the time-periods between each student first and last submission timestamps, we categorize students into three submission groups -- students who complete all submissions in less than five minutes, between five minutes and one hour, and more than one hour. In figure \ref{fig:figure_2} panel (b), we show the number of students in each of the three groups and color the students based on the number of submissions they make. In this MOOC, 66\% of students complete all submissions within five minutes, 10\% of students complete in between five minutes to one hour, and 24\% of students complete in more than one hour. Students who complete all submissions within five minutes typically submit between one and three times and rarely submit four or more times. For students who complete submissions in the other two submission groups -- between five minutes to one hour and in more than one hour -- they are equally likely to make one to four or more than four submissions.

Because the majority of students make multiple submissions within a very short time-period (i.e. less than five minutes), we are motivated to study whether students have made actual code changes between these submissions. We collect information for each student making more than one submissions as shown in the sample data table in figure \ref{fig:figure_1}(a). First, we find student code commit histories for all three required files on the GitHub server. Second, we compare each of the commit timestamp in each of the file found on GitHub to all submission timestamps in MOOC. Third, for each file, we count the number of code commits that happen between student first and last submission timestamps in MOOC. Finally, we add up the code commits in all three required files. If the commit number is 0, it indicates that there is no code change on GitHub in between student first and last submission timestamps in MOOC. Otherwise, it is the number of changes. To further track how many additions and deletions the student makes in one commit, we also scrape the file update page from the commit history on GitHub to retrieve the number of changes to each file that the student pushed to GitHub. We then add up changes in all three files to obtain a total change count. Because the number of changes include both the number of additions and deletions to the files, we use the change count to quantify student code change in our analysis. We then show violin plots for the total number of commits from the three required files in figure \ref{fig:figure_2} panel (c) and the number of changes in panel (d) for the three separate submission groups -- students who complete all submissions in less than five minutes in red color, between five minutes and one hour in green color, and more than one hour in blue color. For illustration purpose, the number of commits and changes in figure \ref{fig:figure_2} (c) and (d) are shown in log2 scale but the actual numbers of commit and change are labeled on the y-axis.

We observe that the 5,803 students in the first submission group (i.e. students complete all submissions within five minutes) rarely commit any changes to their files. The 830 students in the second submission group (i.e. students complete all submissions in between five minutes to one hour) tend to make less than three commits and less than 200 lines of changes to their files but are unlikely to make commits for more than ten times. The 2,108 students in the third group (i.e. students complete all submissions in more than one hour) have approximately the same data distribution as in the second group, but with more proportions of students committing approximately three times and 120 lines of changes. Also, in the third submission group, students make approximately the same number of commits and changes no matter they spend hours or days to complete the last submissions. The number of commits and changes are independent of the time that students spend on re-submissions.

\section{Relationship between MOOC submissions and peer grades}
\label{sec:submission_grade}

\begin{figure}
\centering
\includegraphics[width=\linewidth]{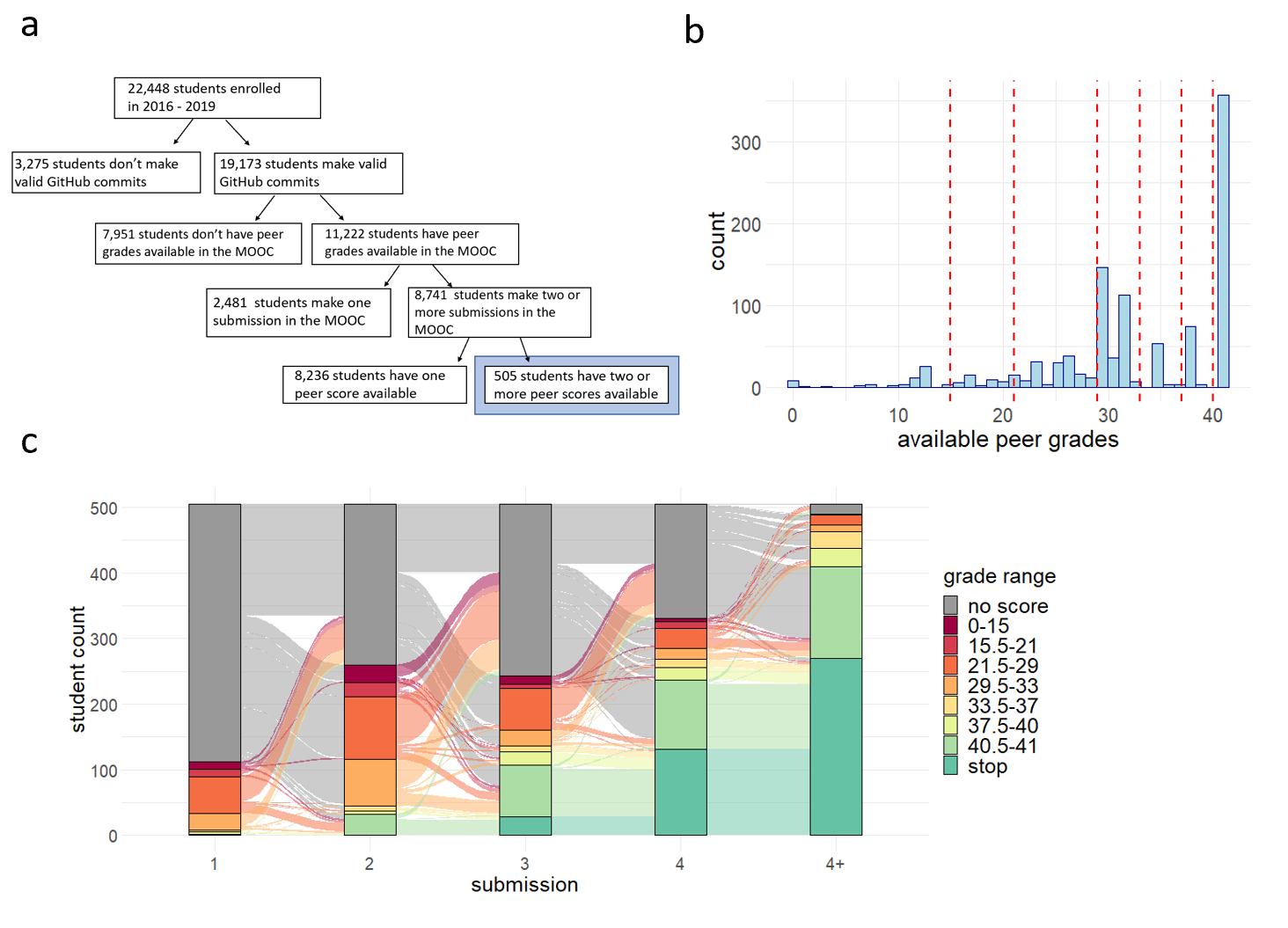}
\caption{\textbf{Student peer grade performances over re-submissions.} We show in panel (a) the data description flowchart with blue color to highlight the desired population in this study, panel (b) the histogram of student available peer grades with dotted vertical lines to highlight grade-range clusters, and panel (c) the alluvial plot to trace student peer grades in grade-range clusters from the first to the last submissions.}
\label{fig:figure_3}
\end{figure}

The grading strategy for this assignment in the MOOC course is relies on  peer-review. The peer-reviews are beneficial for students to practice knowledge learned in class while grading other peer projects. However, peer review in MOOCs may have variable participation\cite{suen2014peer} and scoring may be correlated with student knowledge \cite{meek2017peer}. In the MOOC we are studying, the final project instruction includes the following parts: (a) it requires a student to submit a link to the GitHub repository with the code for performing data analysis in "run\_analysis.R". A student should include a "readme.md" in the repository describing how the R script works and "codebook.md" describing the variables. The grading rubric for part (a) is first for graders to evaluate whether the student submits a valid GitHub repository link with the required scripts that perform the right analysis, and graders can assign grades 0, 6, or 12 for evaluation. Second, graders should evaluate the quality of the code book and readme files, and assign grades 0, 6, or 12 for the code book and 0 or 4 for the readme file. Third, graders should evaluate whether the work submitted for this project is the work of the student who submitted it and assigns 0 or 1 point for evaluation. The grading rubric for part (b) is for graders to evaluate whether the output of the R script is a tidy data set, and assign 0, 6, or 12 for evaluation.

We have described the peer-review grading schema in Section \ref{sec:introduction} that the final peer grade shown in MOOC is an average grade from multiple peer graders. There are some limitations to this peer review scheme. The peer-review grading system allows time to decide which students grade which submissions, and it is not guaranteed that every submission for a student is assigned to peers and gets a peer grade. For example, if a student makes four submissions during a short period, there is not enough time for the MOOC grading system to respond and assign all four submissions to peers for grading. In this case, any of the four submissions can be graded (see an example in the sample data table in figure \ref{fig:figure_1}(a)). Usually, the last submission tends to be assigned with a peer grade, but it is also possible that any of the first three submissions or more than one submissions are assigned with peer grades. A student may receive one or more peer grades when making multiple submissions in the MOOC. Based on the grading policy in the MOOC platform, when a student receives more than one peer grades over submissions, the highest available score is assigned to the student as the final peer grade for this project.

Among the 8,741 students who make two or more submissions in MOOC as described in Section \ref{sec:submission_grade}, 94\% of the population (i.e. 8,236 students) receive only one peer grade, and the remaining 6\% (i.e. 505 students) receive two or more peer grades over submissions. In figure \ref{fig:figure_3} panel (a), we show the new data structure flowchart which is an extension to figure \ref{fig:figure_2}(a) described in Section \ref{sec:submission_commit}. To address our question in finding the relationship between re-submissions and peer grade increases, we focus on the subset of 505 students who receive more than one peer grades, so we can then trace student peer grade performances over submissions. These 505 students are highlighted with blue color in the flowchart shown in figure \ref{fig:figure_3}(a). We also consider the three submission groups described in Section \ref{sec:submission_commit} -- students who complete all submissions in less than five minutes, between five minutes and one hour, and more than one hour. In this study, the entire population (i.e. 505 students) fit into the last submission group -- they complete all submissions in MOOC in more than one hour.

In the final project of the MOOC course, student grades can range from 0 to 41. To study the distribution of student grades among multiple scoring students, we extract all available peer grades for the target population of 505 students. The resulting histogram is shown in figure \ref{fig:figure_3} panel (b). We observe that the density distribution of student peer grades in the course shows peaks near common scores and can be approximated as a Gaussian mixture model \cite{bonett2000sample, reynolds2009gaussian}. Based on this finite normal mixture model, we use mclust \cite{fraley1998mclust, fraley2012mclust}, a model-based clustering algorithm that combines model-based hierarchical clustering \cite{vaithyanathan2013model,fraley1998many}, EM \cite{mclachlan2007algorithm} for mixture estimation, and the Bayesian Information Criterion (BIC) \cite{neath2012bayesian} in comprehensive strategies for clustering, to classify student peer grades. This approach yields seven grade-range clusters -- 0 - 15, 15.5 - 21, 21.5 - 29, 29.5 - 33, 33.5 - 37, 37.5 - 40, 40.5 - 41. We add vertical dotted lines on the histogram in figure \ref{fig:figure_3}(b) to highlight these grade-range clusters.

We then use the clustered grade-range to further evaluate student grade performances over submissions. Among the 505 students, 29 students make two submissions and stop, 103 students make three submissions and stop, 138 students make four submissions and stop, and 235 students make more than four submissions. For each student, we rank all submissions by project submission timestamps in the increasing order and extract the peer grade assigned to each submission. We then classify the peer grades into one of the seven grade-range cluster described above and trace student peer grades in the clustered ranges over submissions using the timestamps in the increasing order. The resulting alluvial plot is shown in figure \ref{fig:figure_3} panel (c). We set the seven grade-range clusters in different colors to better distinguish the changes in student peer grades over submissions. Here dark red color indicates the lowest grade-range cluster (i.e. 0 - 15) and green color that indicates the highest grade-range cluster (i.4. 40.5 - 41). To identify when students complete all of their submissions, we include a "stop" status (in dark green color) in the plot. For example, if a student makes three submissions, the student goes to the "stop" status in the fourth submission. Also, because a student may not receive peer grades for all submissions, We include a "no score" status (in grey color) to label such submissions without grades assigned. In figure \ref{fig:figure_3}(c), we use grade-range clusters instead of actual grades to trace the flow of student grade changes in the first, second, third, fourth, and fifth or more submissions. The submission times are listed as "1", "2", "3", "4", and "4+" on the x-axis in the plot. For students who submit more than four times, we extract the student's highest grade above the fourth submissions and show this highest grade in the "4+" submission category. In this way, the alluvial plot reflects the flow of student grade changes over submissions.

We observe from figure \ref{fig:figure_3}(c) that student peer grades increase on re-submissions on average. In detail, 92\% of the population (i.e. 466 students) obtain grade improvements, 4\% of the population (i.e. 21 students) obtain the same grades, and 4\% of the population (i.e. 18 students) obtain grades that decreases over submissions. Another observation is that students whenever receive "high" grades between 37.5 to 41 tend to stop making more submissions. Students who receive no score or "low" grade such as between 0 to 33 tend to make an additional one to three submissions, but they are also less likely to submit more than four times. For these students, re-submissions can potentially lead to peer grade increases. Overall, student peer grades have substantially improved on average over submissions. The number of students who receive no scores also substantially decreases as students make more submissions.

\section{Relationship between code changes on GitHub and peer grades in MOOC}
\label{sec:commit_grade}

\begin{figure}
\centering
\includegraphics[width=0.9\linewidth]{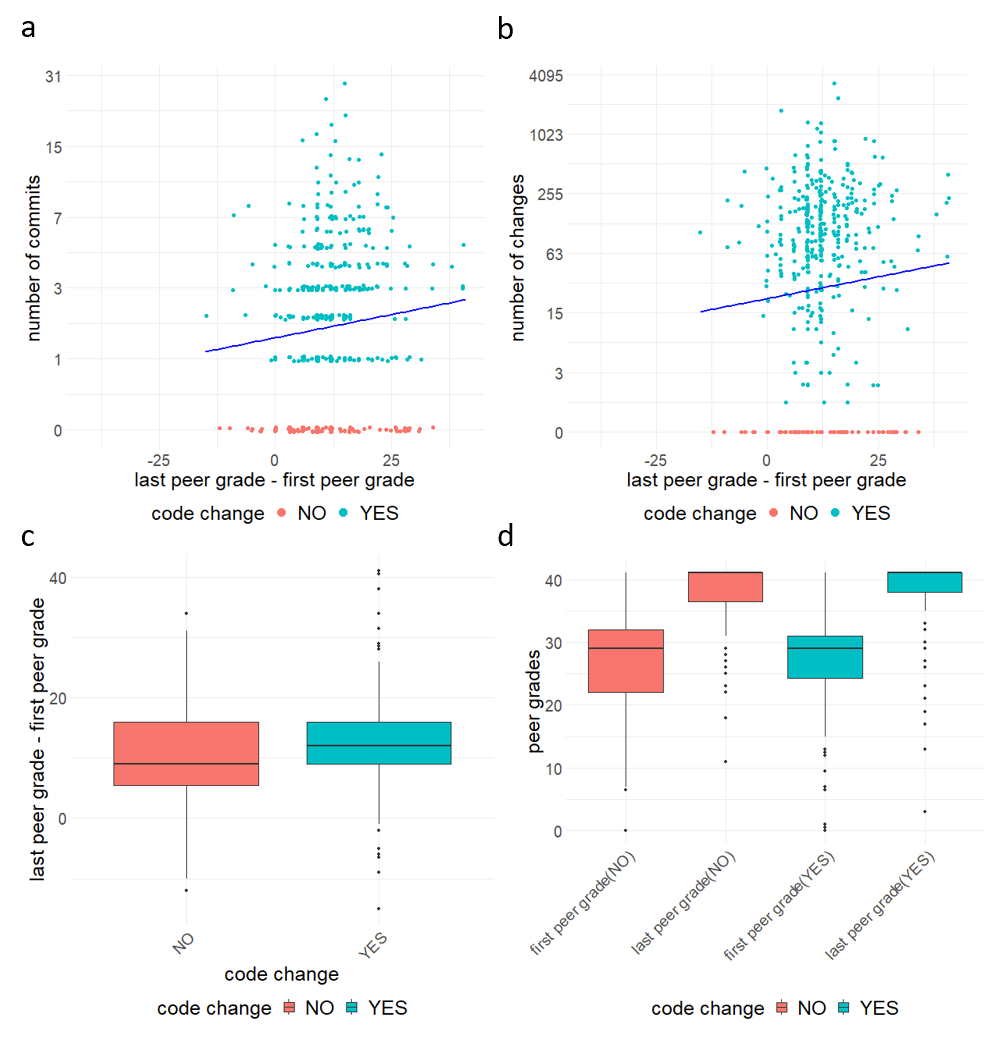}
\caption{\textbf{Student code change on GitHub and peer grades received in MOOC.} On the x-axis in panel (a) and (b) are the grade difference between student last available grades and first available grades. On the y-axis are the number of commits on GitHub in panel (a), and the number of changes on GitHub in panel (b). We show panel (c) the box-plot of student grade differences between the first and last available peer grades in the "NO" and "Yes" groups -- students do not make and make code changes to the files on GitHub, and (d) the box-plot of student first available grades in the "NO" group, student last available grades in the "NO" group, student first available grades in the "YES" group, and student last available grades in the "YES" group.}
\label{fig:figure_4}
\end{figure}

In Section \ref{sec:submission_grade}, we associate student peer grades to submissions in MOOC and observe that student peer grades have substantially increased on re-submissions on average. However, we observe in Section \ref{sec:submission_commit} that there may or may not be any code changes to the final project on GitHub between student submissions in MOOC. So here we aim to further study whether student peer grade increases are related to actual code changes to the files. Again we focus on the population of 505 students as described in figure \ref{fig:figure_3}(a) who make two or more submissions and receive more two or more grades in MOOC.

We have introduced MOOC grading procedure in Section \ref{sec:introduction} and Section \ref{sec:submission_grade}. In short, the MOOC first receives a submission, then arranges multiple peers to grade this submission, and finally assigns the average grade from peer graders to this submission. A grade timestamp will be generated in the MOOC platform when a peer grade is assigned to a submission. We can then use this grade timestamp to keep track of student grades over time.

Among the target population of 505 students who receive more than one peer grade, we first use the grading timestamps to rank student grades in the order of when a student receives them. Second, we use the ordered timestamps to locate the student's first and last available peer grades. Third, we count the number of GitHub commits and changes (i.e. sum of additions and deletions) in the three required files (i.e. "readme.md", "codebook.md", and "run\_analysis.R" ) that have been made between the timestamps of the first and last peer grade. We then associate the grade differences between a student's first and last peer grades to the commit counts and change counts that happen between the grade timestamps.

The resulting relationships between the grade differences and the number of commits and changes are shown as scatter plots in figure \ref{fig:figure_4} panel (a) and (b). We show the number of commits and changes in the log2 scale but label them in actual values on the y-axis. In both cases, we fit a linear regression model \cite{weisberg2005applied} to data with the number of commits/changes (log2-transformed) as the dependent variable and grade differences as the independent variable. The coefficient of the regression model shown in figure \ref{fig:figure_4}(a) is 0.013 (P = 0.03) which means we expect an increase of about 9.4\% in the number of commits for each 10 point score difference. The coefficient of the regression model shown in figure \ref{fig:figure_4}(b) is 0.029 (P=0.10) which means we expect about a 22\% increase in the number of line changes for each 10 point score difference. The results show that there exists a weak relationship between the number of commits and grade differences, and the number of commits slightly increases when student grades increase. we have also calculated the Spearman's correlation \cite{bonett2000sample} between the number of commits and grade differences (r = 0.15) and between the number of changes and grade differences (r = 0.12). Based on these observations, we conclude that there is a positive, but weak relationship between grade changes and code changes to the files on GitHub.

We observe that a group of students who do not make code change also have changes in grades between the grade timestamps (red dots in both figure \ref{fig:figure_4} panel (a) and (b)). We divide the population of 505 students into two groups (i.e., "NO" and "YES"). In the "NO" group, there are 135 students who do not make any code changes to the files on GitHub (in red color). In the "YES" group, there are 370 students who make at least one code commit or change to the files on GitHub (in green color). In figure \ref{fig:figure_4} panel (c), we show a box-plot on grade differences between students' first and last available scores in the "NO" and "YES" groups (figure \ref{fig:figure_4}(c)). To determine whether the means of the two groups are equal to each other, we conduct a t-test \cite{owen1965power} to compare the two groups. The mean grade difference is 11.3 in the "NO" group and 12.4 in the "YES" group while the resulting p-value is 0.25. The results indicate that there are no statistically significant mean differences in grade changes between the "NO" and "YES" groups.

Within the "NO" and "YES" groups, we compare students' first and last available grades and show the results in the box-plot in figure \ref{fig:figure_4} panel (d). In the plot, we observe substantial grade increases from the first to the last available grades in the "No" group (in red color) as well as in the "YES" group (in green color). To further determine whether the means of the first and last available grades are equal to each other within each group, we conduct t-test to compare the first and last grades in each group, respectively. In the "NO" group, the mean first grade is 26.6 and the mean last grade is 37.9. The resulting p-value is less than 0.01. In the "YES" group, the mean first grade is 26.5 and the mean last grade is 38.9. The resulting p-value is also less than 0.01. Thus, the results indicate that there are statistically significant grade increases between students' first and last available grades in both groups. Although we have observed in Section \ref{sec:submission_grade} that student grade increases are highly correlated with the number of submissions in the MOOC, the grade increases are in fact independent of actual code changes made to the files on GitHub.

\section{Discussion}
\label{sec:discuss}

In this paper, we have combined peer-review submission data from a MOOC and publicly available code repositories that form a part of the submission to analyze the dynamics of massive online open peer review. We observe that there are distinct groups of submission behavior based on the frequency that students re-submit their assignments. We have shown that there is only a loose relationship between the number of submissions and code commits. We also observe a similarly weak relationship between code submissions and grades. However, we do observe that peer grades typically increase upon repeated submission and that the variability in re-graded submissions is high. There are some limitations to our analysis, including that the analysis focuses on the data from as single large MOOC, that some of the Github repositories are missing, and that there are a relatively small number of resubmissions to the assessment in question. However, the approach of programatically using open source code repositories as a secondary source of data shows promise for helping us to understand the dynamics of peer grading at a massive scale. The future steps of the paper are first to validate our observation from this exploratory analysis in other MOOCs with peer-graded coding assignments and second to perform experiments to determine if including previous versions of the code, restricting submissions, or altering review criteria will increase the correlation between submitted code and peer assigned grades. While these experiments are beyond the scope of our initial exploratory analysis, we believe that the innovation of combining open-source code repositories and peer grades, combined with the observation that peer grades do not correlate tightly with code commits, suggests an important and yet undiscovered avenue for research in MOOCs.

\bibliographystyle{unsrt}
\bibliography{ref}

\end{document}